\documentclass{aa}  
\usepackage{graphicx}
\usepackage{txfonts}
\usepackage{natbib}
\usepackage{lineno}

\newcommand{\kms}{$\rm km s ^{-1}$}

\newcommand{\sit}{{\it s}-process}
\newcommand{\sdsscemp}{SDSS~J074238.68+470537.0}
\newcommand{\cemps}{CEMP-{\it s}}

\newcommand{\suda}[1]{#1}
\newcommand{\PFR}[1]{#1}

\usepackage[normalem]{ulem}
\nolinenumbers
\begin{document} 

\title{Abundance analysis of three new extremely metal-poor stars, including one CEMP-s star\thanks{Based on observations made with HDS at the Subaru telescope}
}
\titlerunning{EMP stars}

\author{
P.~Fran\c{c}ois \inst{1,2} \and
T.Suda \inst{3,4}\and
S.Wanajo \inst{5}\and
E. Caffau    \inst{6} \and
P.~Bonifacio \inst{6} \and
W. Aoki \inst{7} \and
S. Yamada \inst{8} \and
M. Y. Fujimoto \inst{8}
}

\institute{LIRA, Observatoire de Paris, Universit{\'e} PSL, Sorbonne Universit{\'e}, Universit{\'e} Paris CitC), CY Cergy Paris Universit{\'e}, CNRS,75014,Paris France
\and
UPJV, Universit\'e de Picardie Jules Verne, 33 rue St Leu, 80080 Amiens, France
\and 
Department of Liberal Arts, Tokyo
University of Technology, 5-23-22 Kamata, Ota-ku, Tokyo 144-8535, Japan
\and
Research Center for the Early Universe,
The University of Tokyo, 7-3-1 Hongo, Bunkyo-ku, Tokyo 113-0033,
Japan
\and
Department of Astronomy, Faculty of Science, Tohoku University, Sendai, Miyagi 980-8578, Japan
\and
LIRA, Observatoire de Paris, Universit{\'e} PSL, Sorbonne Universit{\'e}, Universit{\'e} Paris CitC), CY Cergy Paris Universit{\'e}, CNRS,92195 Meudon, France
\and
National Observatory of Japan, Mitaka, Tokyo, Japan
\and
Faculty of Science, Hokkaido University, Kita 10 Nishi 8, Kita-ku, Sapporo, Hokkaido 060-0810, Japan
}

\date{\today; }

  \abstract
{ We report a detailed abundance analysis of three extremely metal-poor (EMP) stars  thanks to spectra obtained with the high-dispersion spectrograph (HDS) at the Subaru telescope.  
}
{
The aim is to observe a sample of EMP candidate stars selected from the  Sloan Digital Sky Survey (SDSS) and determine their  detailed chemical composition.
}
{
High-resolution spectra of three  faint turn-off stars using HDS on the Subaru telescope were acquired.  We used standard 1D models to compute the abundances
 of several elements (Li, C, Mg, Ca, Sr, and Ba). The stars we analysed have a metallicity [Fe/H] of between -3.02 and -3.57 dex. We also measured accurate radial velocities that, coupled with the \textit{Gaia} astrometric data, allowed us to 
 compute actions and Galactic orbits for the stars.
}
{ Our analysis confirms that the  stars are genuine EMP stars. One of them  (SDSS~J074238.68 $+$470537.0) can be classified as a \cemps\ star with the highest [Mg/Fe] ratio ever measured in a turn-off \cemps\ star with [Mg/Fe] = 1.85 dex. For the star SDSS~J012630.30$ +$073029.3 ([Fe/H]= -3.02 dex), we have a clear detection of lithium with A(Li) = 2.05. 
Kinematically, SDSS~J214340.08$-$002835.5 is a thick-disc star, while the other two belong to the halo and are candidate members of the Gaia-Sausage-Enceladus.}
{}

\maketitle
\nolinenumbers
%
\section{Introduction\label{intro}}
The study of the most metal-poor stars is a very active field of research. 
The recent review of \citet{bonifacio_most_2025} covering the last ten years of 
 extremely metal-poor (EMP) studies collects  the results from more than 400 articles published since 2014, a good indicator of the  community's collective effort to better understand the earliest stage of the chemical evolution of the  Galaxy and the nearby Local Group satellite galaxies.
The selection of the EMP candidates relies on a variety of  algorithms applied to large spectroscopic surveys such as  HK \citep{beers_search_1985, beers_search_1992, cayrel_first_2004}, HES \citep{christlieb_finding_2003, christlieb_stellar_2004}, LAMOST  \citep{deng_lamost_2012, li_four-hundred_2022}, DESI \citep{cooper_overview_2023, koposov_desi_2024, allende_prieto_gtc_2023}, SDSS  \citep{york_sloan_2000,ludwig_extremely_2008}, or dedicated photometric surveys  such as   Pristine   \citep{starkenburg_pristine_2017, aguado_pristine_2019}  or Skymapper   \citep{yong_high-resolution_2021}  
with the aim of finding, with the highest level of purity,  a clean selection of EMP stars. An update to the current experiments can be found in \citet{bonifacio_most_2025}. The principle is to select stars in the  temperature  range of F-G stars (typically 4500 to 6500 ~K)  with faint absorption lines. As most of the Fe lines are rather weak, the selection
is  generally based on the Ca H and K lines, the strongest atomic lines available in the spectrum. The detection of the molecular CH G band is also an important element in this search. A complete study of these EMP candidates requires a high-resolution spectroscopic follow-up; as the stars are metal-poor,  the absorption lines of most the elements are rather faint, and a high signal-to-noise ratio (S/N) is  necessary to provide 
the abundances of the largest number of elements. 
These studies revealed an unexpected  diversity of the chemical composition of the EMP stars that can be used to constrain the models of formation and evolution of the Galactic chemical evolution and the nucleosynthesis of the first generations of supernovae.
\PFR{Recent high-resolution spectroscopic studies have further confirmed the diversity of chemical-abundance patterns observed among EMP stars (Bandyopadhyay et al. 2024). Among the EMP stars, a rather large fraction have a high level of carbon abundance, classified as carbon-enhanced metal-poor (CEMP) stars. These CEMP stars   are  divided into several subclasses according to their neutron-capture-element abundances \citep{beers_discovery_2005,aoki_spectroscopic_2007}.
A recent work by \citet{lee_new_2025}  has proposed a new morphological Group IV of CEMP-no stars, which is characterised by high absolute carbon abundances (A(C) > 7.39) at very low metallicity ([Fe/H] $\le$ -3.1) and no neutron-capture enhancements.
}

The stars we analysed in this work were selected from the list of
EMP star candidates \citep{bonifacio_topos_2021} used to derive the metal-weak tail  of the metallicity distribution function of the Milky Way.  The metallicity estimates
of the SDSS spectra are based on the method developed by \citet{ludwig_extremely_2008} essentially confirmed by the analysis at higher resolution. This method has been very successful and has led to a series of papers that have already been published \citep{caffau_extremely_2011, caffau_x-shooter_2011, caffau_primordial_2012, caffau_topos_2016, bonifacio_topos_2015, bonifacio_topos_2018, francois_topos_2018, francois_detailed_2020, suda_detailed_2025} .

The importance of the study of these  metal-poor stars  is that they are formed out of gas that has been very likely enriched by the ejecta of a single or a few supernovae. From the determination of the chemical composition of these stars, we were able to derive important hints regarding the nucleosynthesis in the first generation of stars that polluted the primordial gas and on the chemical  \PFR{ inhomogeneities} during the early evolution of our Galaxy.  Moreover, these abundance determinations can also be used to constrain the scenario of formation of the first low-mass stars \citep{caffau_extremely_2011}.
In this article, we report the detailed analysis of three new extremely metal-poor candidates for which we obtained high-resolution spectra.

\section{Observations} 

The observations were  carried out with the High Dispersion Spectrograph (HDS) installed on the Subaru telescope \citep{noguchi_high_2002}. The wavelength coverage goes from  4084 \AA ~to  6892 \AA  . 
A $2\times 2$ binning  mode was  adopted, resulting in a resolving power of about 40\,000.  
The logbook of the observations is given in Table \ref{tab:obslog}. Standard data-reduction procedures were carried out with the IRAF Echelle package.\footnote{ IRAF is distributed by the National Optical Astronomy Observatories, which is operated by the Association of Universities for Research in Astronomy, Inc. under cooperative agreement with the National Science Foundation.} Particular care was taken to remove the sky background, although most of the observations  were  not affected by the moon illumination.  \

\begin{table*}
 \caption{Observation log. } 
\label{tab:obslog}
\centering
\begin{tabular}{l c c c  }
\hline\hline
  Object          &  g magnitude  &  Observation date &  Exp. time  [s]    \\      
\hline

SDSS~J012630.30$+$073029.3   & 16.02  &         2024-10-30T08:24 &      3600.0  \\  
SDSS~J012630.30$+$073029.3   & --  &            2024-10-30T09:24 &      3600.0  \\         
SDSS~J074238.68$+$470537.0   & 17.70  &         2024-10-30T10:41 &      3600.0  \\  
SDSS~J074238.68$+$470537.0   & --  &            2024-10-30T11:42 &      3600.0  \\         
SDSS~J074238.68$+$470537.0   & --  &            2024-10-30T12:42 &      3600.0  \\         
SDSS~J074238.68$+$470537.0   & --  &            2024-10-30T13:43 &      3600.0  \\
SDSS~J214340.08$-$002835.5   & 17.72  &         2024-09-06T07:39 &      3600.0         \\  
SDSS~J214340.08$-$002835.5   & --  &            2024-09-06T08:40 &      3600.0  \\         
SDSS~J214340.08$-$002835.5   & --  &            2024-10-30T05:12 &      3600.0  \\  
SDSS~J214340.08$-$002835.5   & --  &            2024-10-30T06:13 &      3600.0  \\         
SDSS~J214340.08$-$002835.5   & --  &            2024-10-30T07:14 &      3600.0  \\         

  \hline
  \end{tabular}
   \end{table*}

\section{Analysis}

\subsection{Kinematics} \label{sec:kine}

To investigate the kinematics of these three stars 
we used the positions and proper motions provided
in \textit{Gaia} DR3 \citep{gaiadr3} as input; however, the parallaxes of SDSS~J074238.68 $+$470537.0 and 
SDSS~J214340.08 $-$002835.5 have relative errors that exceed 30\%. For this reason,
we decided to use the photogeometric distances of \citet{bailer-jones_estimating_2021}. 
SDSS~J012630.30 $+$073029.3 has a relative error on the parallax of about 8\% and  the implied
distance is consistent, within errors, with its photogeometric distance, we thus decided to also use
the latter for this star. 
To compute the actions we used the {\tt galpy} python library \citep{bovy_galpy_2015}
and selected the gravitational potential MWPotential2014.
Some of the quantities computed by {\tt galpy} useful for the classification of the stars
are provided in Table\,\ref{Tab:kin}.
We  also used {\tt galpy} to integrate
the orbits of the three stars retroactively for 1\,Gyr.

\begin{table*}
\caption{Some of the kinematical quantities useful for classifying the stars.}
\centering
\label{Tab:kin}
\begin{tabular}{lrrrrrrr}
\hline
  \multicolumn{1}{c}{Object} &
  \multicolumn{1}{c}{$r_{ap}$} &
  \multicolumn{1}{c}{$r_{peri}$} &
  \multicolumn{1}{c}{e} &
  \multicolumn{1}{c}{$z_{max}$} &
  \multicolumn{1}{c}{E} &
  \multicolumn{1}{c}{$L_z$} &
  \multicolumn{1}{c}{$J_r$} \\
 & kpc & kpc  & & kpc & $\rm km^2 s^2$ & $\rm kpc\, km\,s^{-1}$ & $\rm kpc\, km\,s^{-1}$\\ 
\hline
  SDSS J012630.30+073029.3 & 13.4 & 0.1 & 0.99 & 7.4 & $-40846.5$ & $-37.4 $& 929.8\\
  SDSS J074238.68+470537.0 & 14.2 & 0.5 & 0.93 & 2.4 & $-39849.4$ & $249.7 $& 1010.2\\
  SDSS J214340.08-002835.5 & 9.34 & 6.4 & 0.19 & 4.2 & $-42233.2$ & $1397.2$& 38.2\\
\hline\end{tabular}
\end{table*}

\begin{figure}[ht!]
 \begin{center}
\includegraphics[width=7.5cm]{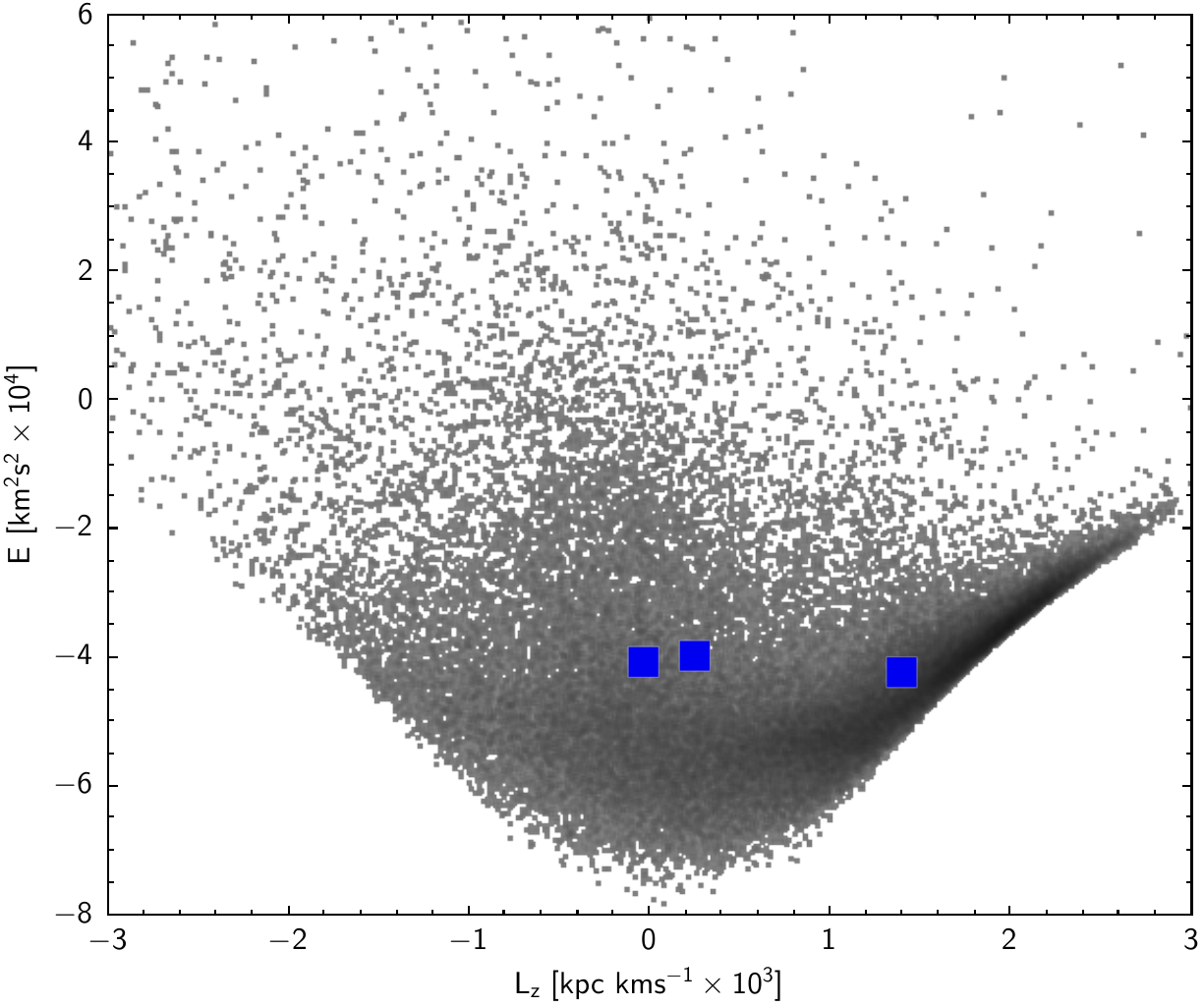} 
 \end{center}
\caption{Our program stars (filled blue squares) in the E versus $L_z$ plane. The grey points in the background
represent the good-parallax sample from \citep{bonifacio_topos_2021}.}
\label{fig:lz_e}
\end{figure}

In Fig.\,\ref{fig:lz_e} we show the angular momentum--energy
diagram for our stars, compared to the good parallax sample
of \citet{bonifacio_topos_2021}.
The stars SDSS~J012630.30$+$073029.3 and  SDSS~J074238.68$+$470537.0 have a clear
halo kinematics, and according to the criteria of 
\citet{feuillet_selecting_2021} they are candidate members
of the Gaia-Sausage-Enceladus \citep{belokurov_co-formation_2018, haywood_disguise_2018, helmi_merger_2018}
halo substructure.
Instead, SDSS~J214340.08$-$002835.5  is a thick-disc star, as is clear from Fig.\,\ref{fig:lz_e}.
Its orbit has a low  eccentricity of 0.19, and it  is confined to 
$\pm 4$\,kpc around the Galactic disc.
The actions that we computed for SDSS~J012630.30$+$073029.3 are  
very close to those provided by \citet{bonifacio_topos_2021}. The main differences
are the radial velocity, which is 85.8 \kms\  from the Subaru spectrum and
68.3\,\kms\  from SDSS, and the distance; we used the photogeometric distance for this work, 
while  \citet{bonifacio_topos_2021} used the \textit{Gaia} DR2 parallax, corrected for a global
zero point of -0.029 mas.

\subsection{Stellar parameters} \label{sec:param}

\begin{table*}
 \caption{Adopted stellar parameters for the list of targets. } 
\label{tab:stellar_parameters}
\centering
\begin{tabular}{l c c c c c c}
\hline\hline
  Object          &  ${\rm T}_{eff}$ & log~g  & [Fe/H]   & $\xi$ (km/s)  &  $v_{r}$ (km/s) & S/N @ 480 nm \\      
\hline

SDSS~J012630.30$+$073029.3   & 6278  & 4.40   &  -3.2  & 1.5  & 85.8 & 30       \\   
SDSS~J074238.68$+$470537.0   & 5907  & 4.43   &  -3.05 & 1.5  &  -163.2& 40      \\  
SDSS~J214340.08$-$002835.5   & 6404  & 4.76   &  -3.9  & 1.5  &  -25.2 &35      \\  

  \hline
  \end{tabular}
  \tablefoot{The last column gives the measured radial velocity of the stars after correction from the barycentric velocity.}
   \end{table*}

The effective temperatures in Table \ref{tab:stellar_parameters} were computed by \citet{caffau_topos_2013}. 
The effective temperature is based on  the photometry, using the 
$(g-z)_{0}$ colour and the calibration described in \citet{ludwig_extremely_2008} and taking into account the 
reddening according to the \citet{schlegel_maps_1998} extinction maps and corrected
as in \citet{bonifacio_search_2000}. As discussed in \citet{bonifacio_chemical_2012}, this selects the stars of the halo turn-off and excludes the majority of white-dwarf stars. 
 
 A microturbulent velocity of 1.5 km s$^{-1}$, appropriate for stars with $\log g \simeq 4$,
 was adopted following the prescriptions of \citet{barklem_hamburgeso_2005}.  The metallicities shown in Table \ref{tab:stellar_parameters}  were computed by \citet{caffau_topos_2013}  using the code MyGisFoS \citep{sbordone_mygisfos_2014} to analyse the  spectra of the stars SDSS~J012630.30 $+$073029.3, SDSS~J074238.68 $+$470537.0, and SDSS~J214340.08 $-$002835.5.
 The surface gravities were computed by \citet{bonifacio_topos_2021} using the above effective temperature, the
 \textit{Gaia} DR3 parallaxes, and the Stefan--Boltzmann equation, assuming a mass of 0.8 $M_\odot$.  
 These results can also be found in \citet{bonifacio_topos_2021}.\footnote{ http://vizier.cds.unistra.fr/viz-bin/VizieR?-source=J/A+A/651/A79}

\subsection{Abundances}\label{secabbo}

We carried out a classical 1D local thermodynamic equilibrium (LTE) analysis using OSMARCS model atmospheres  \citep{gustafsson_grid_1975,   gustafsson_grid_2003,   gustafsson_grid_2008,  plez_spherical_1992, edvardsson_chemical_1993}. The abundances used in the model atmospheres were solar-scaled with respect to the \citet{grevesse_abundances_2000} solar abundances, except for the $\alpha$ elements that are enhanced by 0.4 dex. Corrections on the resulting abundances were considered in order to take into account the difference  between  \citet{grevesse_abundances_2000} and  \citet{caffau_solar_2010} and the \citet{lodders_abundances_2009} solar abundances.
 The solar abundances we adopted are reported in Table\,\ref{tab:solarabbo}.

\begin{table}
\caption{Solar abundances used throughout this paper.}
\label{tab:solarabbo}
\centering
\begin{tabular}{ll}
\hline
Element & A(X)  \\
\hline
C & 8.50  \\
Na & 6.31 \\
Mg  & 7.54  \\
Ca & 6.33 \\
Fe & 7.52 \\
Sr & 2.92 \\
\PFR Y  &  \PFR{2.21} \\
Ba & 2.17  \\
\PFR{Ce} & \PFR{1.58} \\
\PFR{Nd} & \PFR{1.45} \\
\PFR{Eu} & \PFR{0.52} \\

\hline
\end{tabular}
\end{table}

The abundance analysis was performed using the radiative-transfer code {\tt Turbospectrum}, a   LTE spectral -line  analysis  code  developed  by  \citet{alvarez_near-infrared_1998} and \citet{plez_turbospectrum_2012}, which treats scattering in detail. The carbon abundance was determined by fitting the CH band near to 430 nm (G band). The molecular data that correspond to the CH band are described in \citet{masseron_ch_2014}.
 The abundances were determined by matching a synthetic spectrum centred on each line of interest to the observed spectrum.  Table \ref{tab:linelist} gathers the list of lines that were used to measure the abundances or evaluate upper limits in our sample of stars.

\begin{table}
 \caption{Absorption lines used to determine the abundances.} 
\label{tab:linelist}
\centering
\begin{tabular}{l r r r }
\hline\hline
 Element &  Wavelength ({\rm$\AA$})   & \rm{$\chi_{esc}$} & log~gf \\
 
 \hline
 LiI         &   6707.761   & 0.00   &  -0.009  \\
 LiI         &   6707.912   & 0.00   &  -0.309  \\
 CH band     &   4315       &        &          \\
 CH band     &   4324       &        &          \\
 Na          &   5895.92    & 0.00   & -0.190   \\
 MgI         &   5172.698   & 2.71   &  -0.38   \\
 MgI         &   5183.619   & 2.72   &  -0.16   \\
CaI          &   4226.740   & 0.00   &  +0.24   \\   
FeI          &   4202.040   & 1.48   &  -0.70   \\
FeI          &   4260.486   & 2.40   &  -0.02   \\
FeI          &   4271.164   & 2.45   &  -0.35   \\
FeI          &   4325.775   & 1.61   &  -0.01   \\
FeI          &   4383.557   & 1.48   &   0.20   \\
FeI          &   4404.761   & 1.56   &  -0.14   \\
FeI          &   4415.135   & 1.61   &  -0.61   \\
FeI          &   5269.550   & 0.86   &  -1.32   \\
SrII         &   4215.520   & 0.00   &  -0.17   \\  
\PFR{YII}          &   \PFR{4883.684}   & \PFR{1.08}   &   \PFR{0.07}   \\   
BaII         &   4554.036   & 0.00   &   0.16   \\
\PFR {LaII}         & \PFR{4123.218}   & \PFR{0.32}   & \PFR{0.11}   \\ 
\PFR {CeII}         &  \PFR{4137.645}   & \PFR{0.52}  & \PFR{0.44}   \\  
\PFR {NdII}         &  \PFR{4462.979}   & \PFR{0.56}   & \PFR{0.04}   \\  
\PFR {EuII}         &  \PFR{4129.725}   & \PFR{0.00}   & \PFR{0.22}    \\   
   \hline
  \end{tabular}
   \end{table}

\subsection{Errors }

\begin{table}
 \caption{  \PFR{
 Sensitivity of  [X/Fe]  abundances on atmospheric parameters for the stars  SDSS~J012630.30$+$073029.3.}} 
\label{tab:errors}
\centering
\begin{tabular}{l c c c }
\hline\hline
  [X/Fe]            &   $\Delta T_{eff}$ =  100~K  & $\Delta$ log~g = 0.5~dex   &$\Delta$  $ v_{t}$ =   0.5~km/s \\     
\hline
C    &   0.2  &   0.2  &     0.1    \\
Na   &  0.1 & 0.1 &  0.15 \\
Mg &   0.1  &  0.15 &  0.15  \\
Ca &   0.1  &  0.1  &  0.15 \\  
Sr &   0.1  &  0.25  &  0.25  \\
Ba &   0.1  &  0.25   &  0.3  \\
  \hline
  \end{tabular}
  \tablefoot{The other stars give similar results.}
   \end{table}

Table \ref{tab:errors} lists the computed errors in the elemental-abundance ratios due to typical uncertainties in the stellar parameters. The errors were estimated varying  $T_{eff}$ by $\pm$ 100~K, log~g  by $\pm$  0.5~dex, and $ v_{t}$  by $\pm$ 0.5 dex in the model atmosphere of SDSS~J012630.30$+$073029.3   other stars give similar results. 
In this star, we were able to measure the Li, Na, Mg,  Ca, and Sr and set limits for C and Ba  abundances.  The main uncertainty comes from the error in the placement of the continuum when the synthetic line profiles are matched to the observed spectra. In particular, residuals from the  sky subtraction may lead to a decrease of the S/N. As the final spectra were constructed by combining  several exposures obtained at different epochs, and hence different barycentric velocities, the features  of sky residuals are smoothed and may degrade the S/N of the spectra. 
This error is of the order of 0.1 to 0.2,  depending on the S/N ratio of the spectrum and   the chemical species under consideration, the largest value being for the neutron-capture elements.  When several lines are available, the typical line-to-line scatter for a given element is 0.1 to 0.2 dex.

\begin{table}
 \caption{  \PFR{ 
 Estimates of the NLTE corrections for the stars of the sample. } } 
\label{tab:nlte}
\centering
\begin{tabular}{l c c c }
\hline\hline
 [X/Fe] &  SDSS~    & SDSS~  & SDSS~ \\  
 &   J012630.30        &  J074238.68   &J214340.08   \\
&  $+$073029.3        &    $+$470537.0                &      $-$002835.5   \\                 

\hline
Na   &  -0.2 & -0.2 &  -0.2 \\
Mg &   0.15  &  0.1 &  0.15  \\
Sr &   0.2  &  0.10  &  0.2  \\
Ba &   0.1  &  0.1   &  0.1  \\
  \hline
  \end{tabular}
   \end{table}

\subsection{NLTE effects }

\PFR{ Na, Mg, Sr, and Ba are known to be sensitive to departures from LTE (or NLTE effects), in particular in metal-poor stars. To estimate the intensity of this effect, we used the NLTE abundance corrections computed with 1D and 3D NLTE codes 
presented in Figs. 33 and 34 of \citet{bergemann20263dnonlteradiationtransfer}.  The results,  gathered in Table \ref{tab:nlte}, show that the NTLE corrections range from -0.2 to 0.2 dex. }

\begin{table}[ht!]
 \caption{Lithium and  carbon abundances. $\alpha$ and  neutron-capture element  abundance ratios.}   
\label{tab:ratios_2}
\centering
\begin{tabular}{l r r r }
\hline\hline
   Element        &  SDSS    &    SDSS~  &       SDSS~ \\
          &  J012630.30    &  J074238.68   &      J214340.08 \\
          &  $+$073029.3    &  $+$470537.0   &     $-$002835.5 \\
   
 \hline    
 {[Fe/H] }   &        -3.02 &       -3.57   &       -3.07  \\
 A(Li)       &        2.05 & $\le$  2.00    & $\le$  2.20  \\
 A(C)        & $\le$  7.0  &        7.5     & $\le$  8.0   \\
{[Na/Fe] }  &        0.21 &        1.5     & $\le$ -0.24  \\
{[Mg/Fe] }  &        0.30 &        1.85    &        0.34  \\
{[Ca/Fe] }  &        0.49 &        0.54    &        0.34  \\
{[Sr/Fe]}    &       -0.20 & $\le$  0.35    &        0.3   \\
\PFR{[Y/Fe] }    & \PFR$\le$  0.91 & \PFR $\le$  1.86    &\PFR $\le$  1.26  \\
{[Ba/Fe]}    & $\le$ -0.15 &        1.37    & $\le$  0.3   \\
\PFR{[La/Fe] }   & \PFR$\le$  1.99 &\PFR $ \le$  2.84    &\PFR $\le$  2.94  \\
\PFR{[Ce/Fe]}    & \PFR$\le$  2.24 &\PFR $\le$  2.79    &\PFR $\le$  2.69  \\
\PFR{[Nd/Fe]}    & \PFR$\le$  2.27 &\PFR $\le$  3.22    &\PFR $\le$  2.32  \\
\PFR{[Eu/Fe]}    & \PFR$\le$  2.00 &\PFR $\le$  2.75    &\PFR $\le$  3.05  \\   
 \hline

\end{tabular}
\tablefoot{The  abundances are given  as A(X) for Li and C  and  [X/H] for Fe
 and [X/Fe] for Na,  Mg, Ca, Sr, Y, Ba, La, Ce, Nd, and Eu}
\end{table}
{}

\section{Results and discussion}

\subsection{Lithium}

\begin{figure}[ht!]
 \begin{center}
\includegraphics[width=7.5cm]{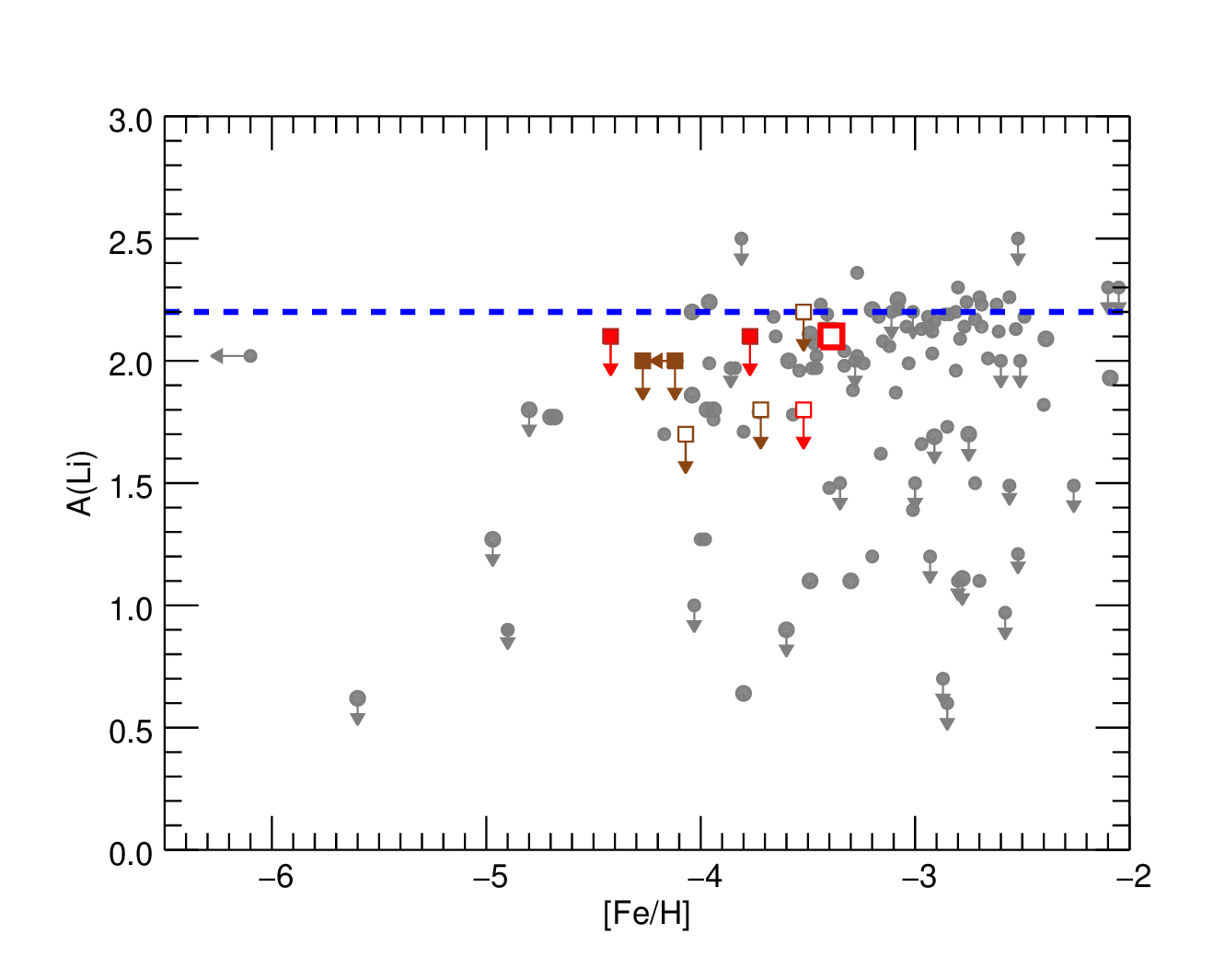} 
 \end{center}
\caption{Lithium abundance in unevolved EMP stars. \PFR{ The filled grey circles refer to literature data}.   Upper limits of the programme stars are shown in red.  Open red symbols represent the  stars  we found with low [n-capture/Fe] upper-limit abundances.  The dashed blue line represents the Spite Plateau as determined
 by \citet{sbordone_metal-poor_2010}.  Details about literature data  can be found in \citet{bonifacio_topos_2018}. 
 Results from our previous analyses \citep{francois_detailed_2020,suda_detailed_2025} of Subaru HDS spectra are shown in brown (open symbols represent stars with low [n-capture/Fe]  upper-limit abundances).  The star SDSSJ002314.00$+$030758.07 with [Fe/H] $<$ -6.10 dex  from \citet{aguado_back_2019} was added.}\label{fig:lithium}
\end{figure}

\begin{figure}[h!]
 \begin{center}
\includegraphics[width=7.5cm]{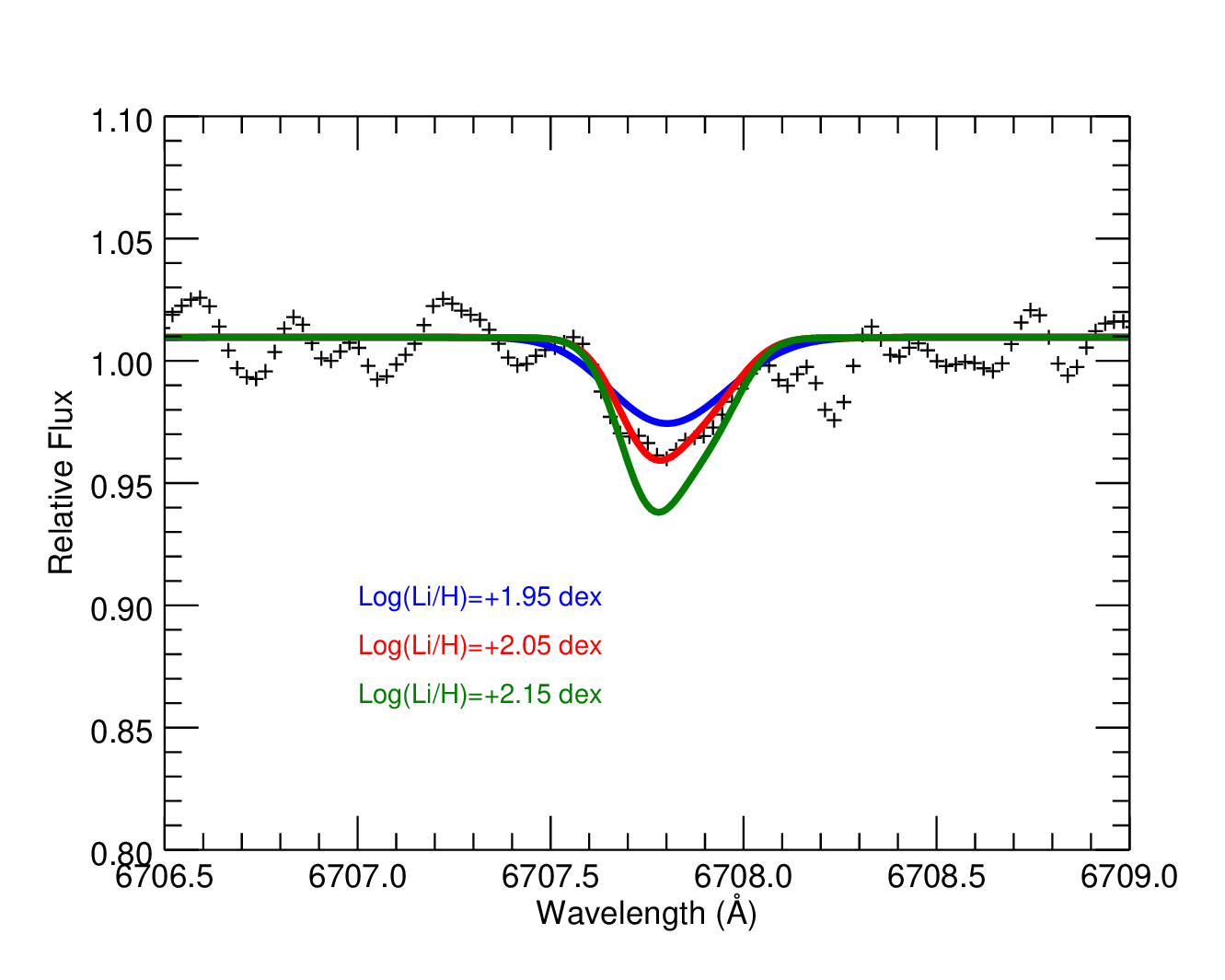} 
 \end{center}
\caption{Comparison of observed spectrum of SDSS~J012630.30$+$073029.3 with  synthetic spectra assuming several Li abundances. The observed spectrum is shown with + signs. The synthetic spectra are represented as lines. The best fit with A(Li)=2.05 is shown in red.  }\label{fig:lithium_fit}
\end{figure}

 Figure \ref{fig:lithium} shows our lithium-abundance results represented as red squares. Open symbols represent stars with low [n-capture/Fe], where [Sr/Fe] or [Ba/Fe] is likely below 1 dex based on upper limits.
 Results from our previous analyses \citep{francois_detailed_2020,suda_detailed_2025} of Subaru HDS spectra are shown as brown symbols.
 Open squares are used for our data because the abundances of neutron-capture elements are not well constrained.
As lithium is diluted in the convective envelopes of red giants, we made the comparison with the lithium abundances determined in unevolved stars \PFR{(shown as grey symbols)} from the literature using the  Stellar Abundances for Galactic Archaeology  (SAGA) database \citep{Suda2008}.
The selection criteria include [Fe/H] $< -2$ and $\log g \geq 3.5$, following the stellar evolution models of 0.8 $M_{\odot}$ at very low metallicity \citep{suda_stellar_2011}. If lithium abundance was reported in multiple papers, we arbitrarily selected one data point, considering factors such as spectral resolution, abundance constraints, the number of measured elements, and other relevant criteria
\citep{frebel_he_2008,matas_pinto_metal-poor_2021,plez_analysis_2005,hansen_first_2011,siqueira_mello_first_2013,lai_detailed_2008,caffau_topos_2016,caffau_extremely_2011,li_enormous_2018,li_four-hundred_2022,thompson_cs_2008,roederer_search_2014,jeong_search_2023,spite_lithium_2015,spite_detailed_2022,bonifacio_topos_2015,bonifacio_topos_2018,lucatello_stellar_2003,hansen_elemental_2015,masseron_lithium_2012,matsuno_high-resolution_2017,  matsuno_lithium_2017,sivarani_first_2004,sivarani_first_2006,placco_g64-12_2016,aoki_carbon-enhanced_2008,aoki_extreme_2010}.
We obtained a detection of lithium in the star SDSS~J012630.30$+$073029.3  
with a value of A(Li) = 2.05 dex. 
The comparison of the observed spectrum with the synthetic spectrum is shown in Fig. \ref{fig:lithium_fit}.
In our sample of three EMP stars, we evaluated the upper limit in two of
them and were able to measure the lithium abundance in the last one (SDSS~J012630.30$+$073029.3).

The star SDSS~J012630.30$+$073029.3 has a lithium abundance  slightly lower\PFR{ (A(Li)}  = 2.05)) than the Spite plateau value \citep{spite_lithium_1982}. This star lies in  the region where the `meltdown' of the lithium plateau is present \citep{sbordone_metal-poor_2010,aoki_lithium_2009,bonifacio_first_2007}.
For two of the remaining  stars, we found  upper limits for the lithium abundance at the level of A(Li) $\simeq$ 2.0 to 2.2 , which is not enough to confirm the meltdown.

\subsection{Carbon}

\begin{figure}[h!]
 \begin{center}
\includegraphics[width=7.5cm]{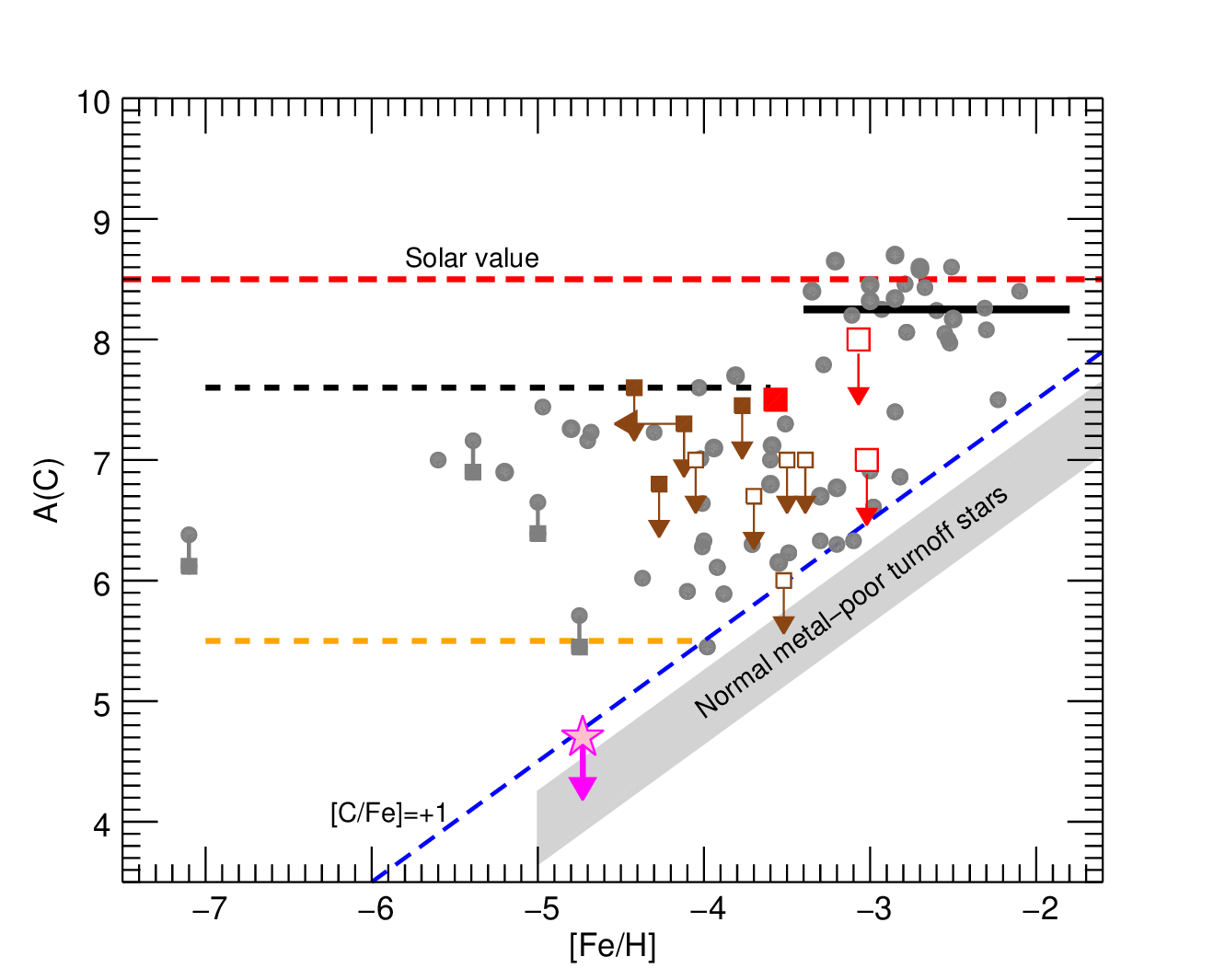} 
 \end{center}
\caption{Carbon abundances A(C) of CEMP stars as a function of [Fe/H]. Large filled red squares represent our results. Down-pointing arrows indicate that the abundances are given as upper limits. Open red symbols represent the  stars for which we found low [n-capture/Fe] upper-limit abundances.
Filled grey circles show data from the literature 
 \citep{sivarani_first_2006, plez_analysis_2005, plez_chemical_2005, frebel_nucleosynthetic_2005, frebel_bright_2006, thompson_cs_2008, aoki_carbon-enhanced_2008, behara_three_2010, masseron_holistic_2010, yong_most_2013, cohen_normal_2013, li_spectroscopic_2015, bonifacio_topos_2018}.   Filled grey squares  show carbon abundances for the same stars  taking into account the  3D corrections from \citet{gallagher_-depth_2016}.    Results from  previous analyses  based on  Subaru HDS spectra  \citep{francois_detailed_2020, suda_detailed_2025} are shown as  brown squares, with open squares being for low-n-capture abundance stars.
       The pink symbol represents SDSS~J102915$+$17292, the normal carbon ultra metal-poor star discovered by \citet{caffau_extremely_2011}.
      The dashed black and orange lines  delimit the low-carbon band. Details can be found in \citet{bonifacio_topos_2018}.}
       \label{fig:carbon_fig}
\end{figure}

In Fig. \ref{fig:carbon_fig} we present our results for the carbon abundance. Our stars are
shown  as red squares. For two stars,  the downward-pointing arrows indicate upper-limit abundances. Open red symbols represent the stars for which we find
low [n-capture/Fe] upper-limit abundances. Literature data were added  \citep{sivarani_first_2006, plez_analysis_2005, plez_chemical_2005, frebel_nucleosynthetic_2005, frebel_bright_2006, thompson_cs_2008, aoki_carbon-enhanced_2008, behara_three_2010, masseron_holistic_2010, yong_most_2013, cohen_normal_2013, li_spectroscopic_2015, bonifacio_topos_2018}. 
Results from our previous analyses \citep{francois_detailed_2020, suda_detailed_2025} of Subaru HDS spectra are shown as  brown squares, with open squares showing low-n-capture abundance stars.
The pink symbol represents SDSS~J102915$+$17292, i.e. the normal carbon ultra metal-poor star discovered by \citet{caffau_extremely_2011}.

For two stars, we could only derive upper limits of the carbon abundance. Given that the S/N is rather low in the region of the G band, 
we were only able to obtain fairly high upper limits, such that we cannot conclude if these stars are CEMP or C-normal.
For the star SDSS~J074238.68$+$470537.0, we measured the carbon abundance and derive an abundance of \PFR{  A(C) } = 7.5, thus  \PFR{ classifying this star as a high-carbon-band CEMP star \citep{yoon_observational_2016,bonifacio_topos_2018}. }

\subsection{$\alpha$-elements and neutron-capture elements}

\begin{figure}[h!]
 \begin{center}
\includegraphics[width=10cm]{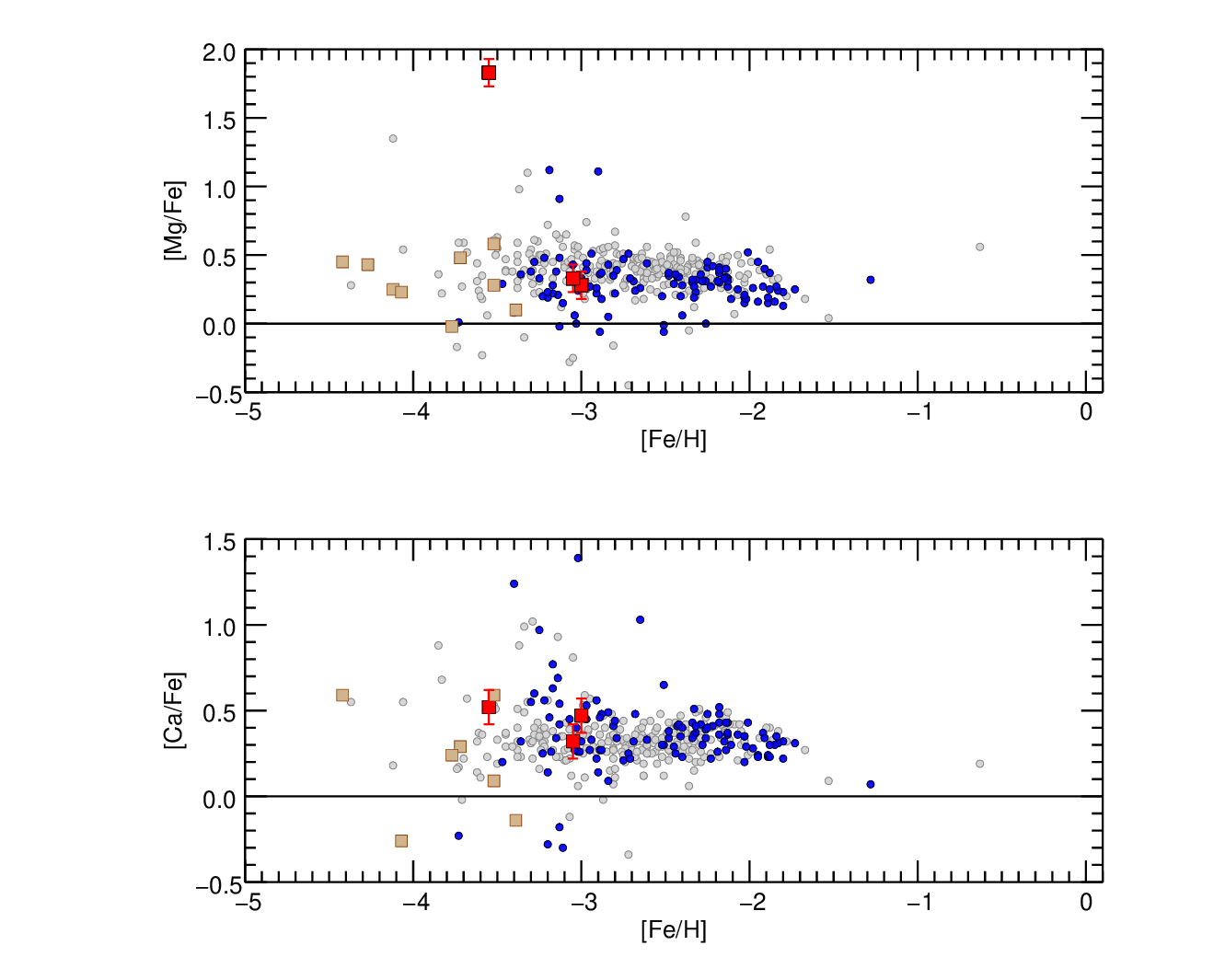} 
 \end{center}
\caption{[Mg/Fe] and [Ca/Fe] abundance ratios as a function of [Fe/H]. Red squares: This paper.   Grey circles:  Evolved stars from \citet{li_four-hundred_2022}. Blue circles: Main-sequence stars from \citet{li_four-hundred_2022}. Light brown squares: Turn-off stars from \citet{francois_detailed_2020} and \citet{suda_detailed_2025} based on HDS spectra.}
\label{fig:alpha}
\end{figure}

In Fig. \ref{fig:alpha},  we present our results for Mg and Ca, the two $\alpha$-elements that can be measured in our spectra, as red squares. Literature data from unevolved  and main-sequence stars from \citet{roederer_search_2014} were added. We chose their data as they represent  a  homogeneous reference sample. 

Two stars of our sample show an over-solar  [Mg/Fe] ratio typical of the vast majority of halo stars. One of the stars (SDSS~J074238.68$+$470537.0) shows an  exceptionally high overabundance of magnesium with a ratio of [Mg/Fe] = + 1.85 dex. This star has also been found to be carbon rich. We discuss this star in more detail later in this paper.

For the three stars of the sample, we find [Ca/Fe] values consistent with what is observed in halo stars, i.e.  an overabundance of Ca with a [Ca/Fe] ratio ranging from +0.3 to 0.5 dex.  However, it is known that a fraction of EMP stars have low [$\alpha$/Fe] ratios. If we combine our results with previous analyses based on HDS spectra \citep{francois_detailed_2020, suda_detailed_2025}, we have a wider [Ca/Fe] ranging from $\simeq -0.25$ ~dex  to $\simeq +0.6$ ~dex,  illustrating the abundance inhomogeneities  operating during the early phases of Galactic chemical evolution as already suggested by \citet{bonifacio_topos_2018}.  It is interesting to note that low [$\alpha$/Fe] ratios are present in turn-off stars, indicating that these abundance anomalies were likely  present when these stars formed.

\begin{figure}[ht!]
 \begin{center}

\includegraphics[width=10cm]{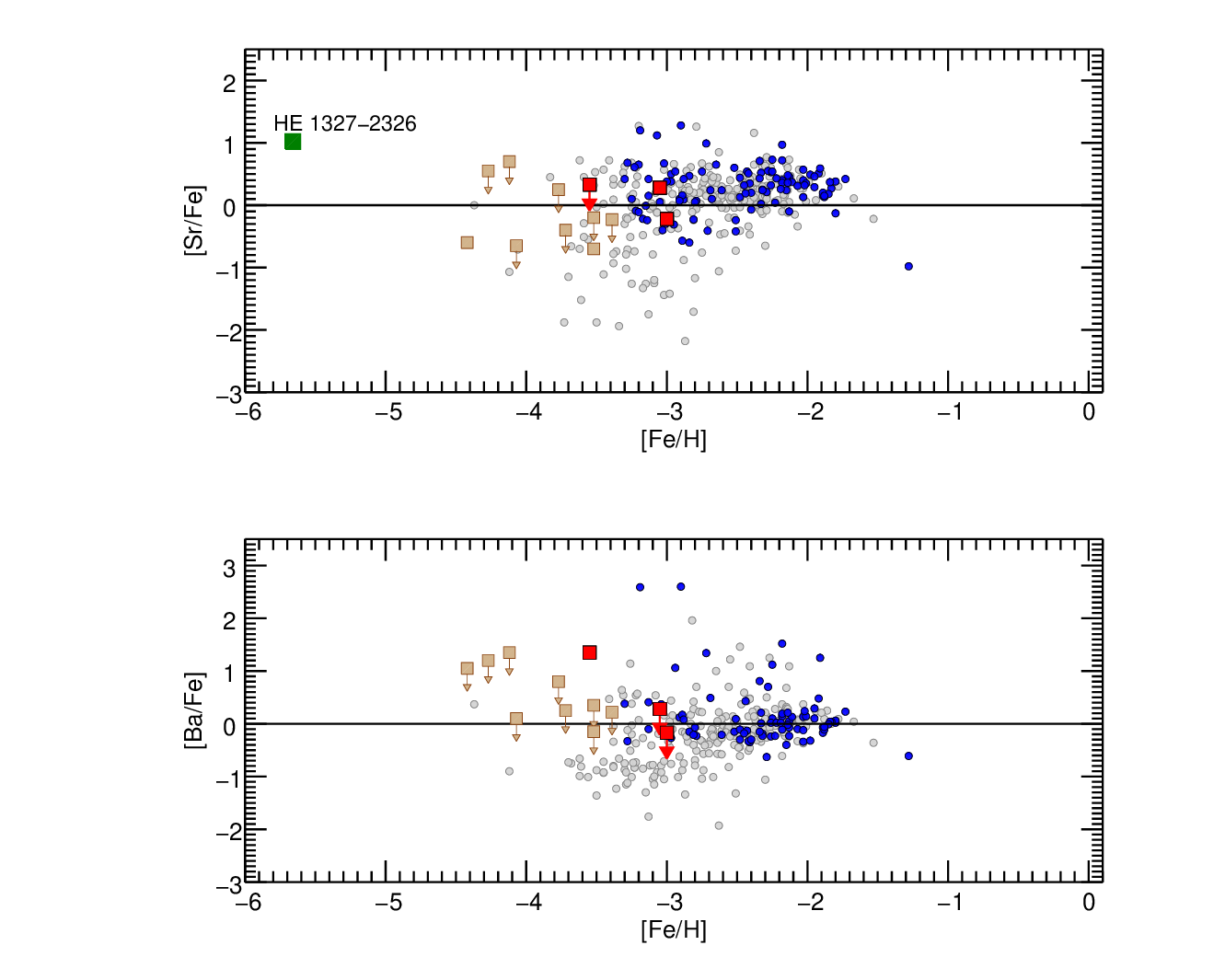} 
 \end{center}
\caption{[Sr/Fe] and [Ba/Fe] as a function of  [Fe/H]. Red squares: This paper. 
       Grey circles:  Evolved stars from \citet{li_four-hundred_2022}. Blue circles: Main-sequence stars from \citet{li_four-hundred_2022}. Light brown squares: Turn-off stars from \citet{francois_detailed_2020} and \citet{suda_detailed_2025} based on HDS spectra. The green rectangle represents HE1327-2326,  a star with an exceptionally high [Sr/Fe] ratio \citep{frebel_he_2008, aoki_he_2006}.}
\label{fig:n_capture}
\end{figure}

Figure \ref{fig:n_capture} shows the  abundance ratios  of the two neutron capture elements  strontium and barium as a function metallicity. Results from 
\citet{li_four-hundred_2022} have been added as representative of a large homogeneous sample of evolved and main-sequence stars. For two stars, we measured the abundance of strontium.

In Fig. \ref{fig:n_capture} we plot the abundance ratios and upper limits of [Sr/Fe] and [Ba/Fe] as a function of [Fe/H]. We also added the results  of \citet{li_four-hundred_2022} from a rather large sample of evolved and main-sequence stars analysed in a homogeneous way. 
Our results are in agreement with the general trend found for the variation of the [Sr/Fe] as a function [Fe/H]. It is interesting to note that the very low [Sr/Fe] ratios
measured in EMP stars are preferentially found in evolved stars. For main-sequence and turn-off stars, the [Sr/Fe] is around the solar value.

For barium, we  determined  upper limits in two stars. The third star, SDSS~J074238.68$+$470537.0,  shows not only a very high [Ba/Fe] ratio of +1.37 but also a high [Mg/Fe] ratio. 
In Fig. \ref{fig:n_capture}, low abundances or undetermined values of neutron-capture elements are indicated by open symbols, as shown in Fig. \ref{fig:carbon_fig}.

In two  stars, we have a detection of Sr with [Sr/Fe] $= -0.20$ dex  and [Sr/Fe] $= +0.30$ dex, similar to the values found in main-sequence stars from \citet{li_four-hundred_2022}.  The results of that work combined with our results and those of our previous studies \citep{francois_detailed_2020, suda_detailed_2025} seem to show that  main-sequence stars exhibit an increasing dispersion as metallicity decreases. The results from \citet{li_four-hundred_2022} show an even higher dispersion for giant stars.
We could measure the barium abundance in the most metal-poor star of our sample, SDSS~J074238.68$+$470537.0, and set upper limits for the other two stars. 
Increasing the sample of  Sr and Ba measurements in main-sequence stars in the metallicity range of $-2.5$ to $-3.0$ would be useful to estimate this dispersion and compare it with  what is  found for the giant sample. At lower metallicity, the measurement of their abundance remains challenging, the exception being the  \cemps\ stars.

\section{SDSS~J074238.68$+$470537.0}

\begin{figure}[ht!]
 \begin{center}
\includegraphics[width=8.5cm]{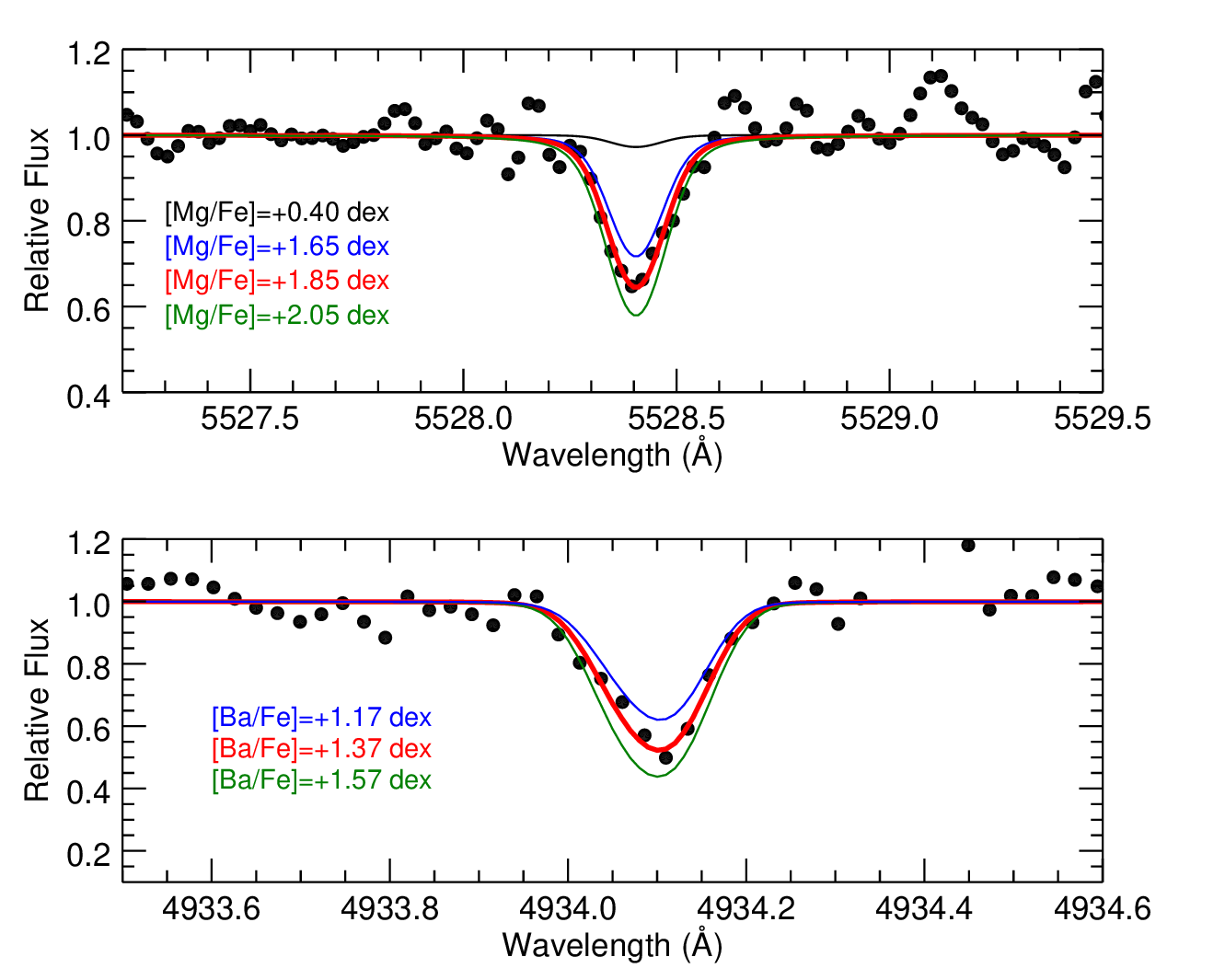} 
 \end{center}
\caption{ Comparison of the observed spectrum of SDSS~J074238.68$+$470537.0  with  synthetic spectra in the region of the  552.8 nm Mg  line and the 493.4 nm Ba line. The observed spectrum is shown with black circles. The synthetic spectra are represented as solid lines.}
\label{fig:MgBa_fit}
\end{figure}

In Fig. \ref{fig:MgBa_fit} we show the comparison of the synthetic spectrum of  SDSS~J074238.68$+$470537.0 in the region of the 552.8 nm Mg  line and the 493.4 Ba line assuming several abundances of Mg and Ba. Thanks to the good quality of the spectrum, we can unambiguously show that this star has high [Mg/Fe] and [Ba/Fe] ratios with respective values of +1.85 dex and 1.37 dex.
In \citet{bonifacio_topos_2018} another Mg-rich star was analysed (SDSS\,J134922.91+140736.9), but the enhancement in Mg was not so extreme ([Mg/Fe]=1.42).

The Na abundance of SDSS~J074238.68$+$470537.0 was measured from  one of the \ion{Na}{I}  D lines, the one at 589.59 nm.
The \ion{Na}{i} lines are known to be severely affected by departures from LTE. We adopted the NLTE correction of 0.2 dex estimated by \citet{takeda_non-lte_2003}. A large overabundance of Na is found in SDSS~J074238.68$+$470537.0 ([Na/Fe]B = + 1.5 ), even after the NLTE correction, leading to a value of [Na/Fe] = +1.3 dex.

\begin{figure}[ht!]
 \begin{center}
\includegraphics[width=9.5cm]{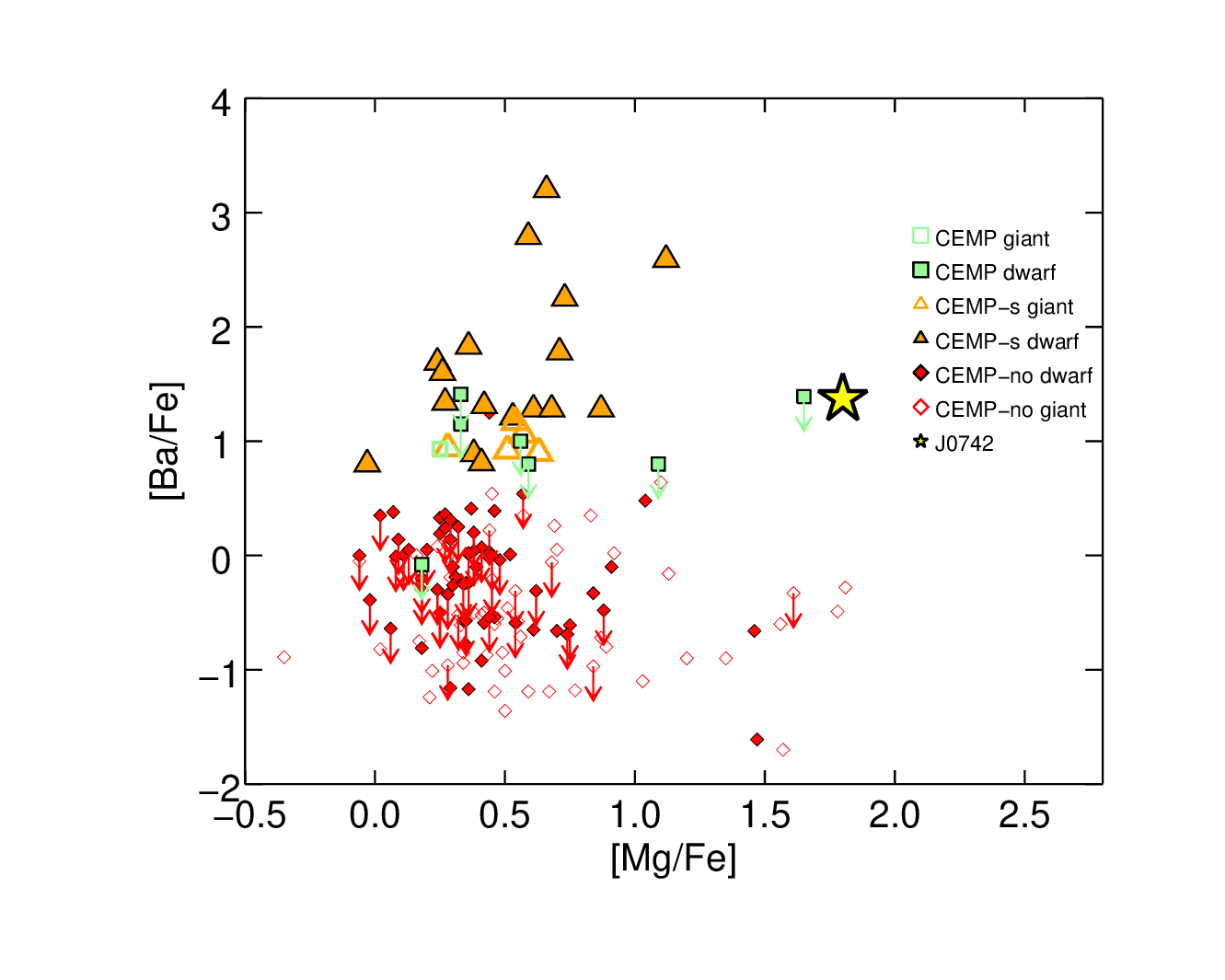} 
 \end{center}
\caption{[Ba/Fe] as a function of [Mg/Fe] for the CEMP stars. Literature data are from the SAGA database. 
Our  newly discovered \cemps\ star SDSS~J074238.68+470537.0 is represented with  a large yellow star symbol. }
\label{fig:BaMg}
\end{figure}

\begin{figure}[ht!]
 \begin{center}
\includegraphics[width=7.5cm]{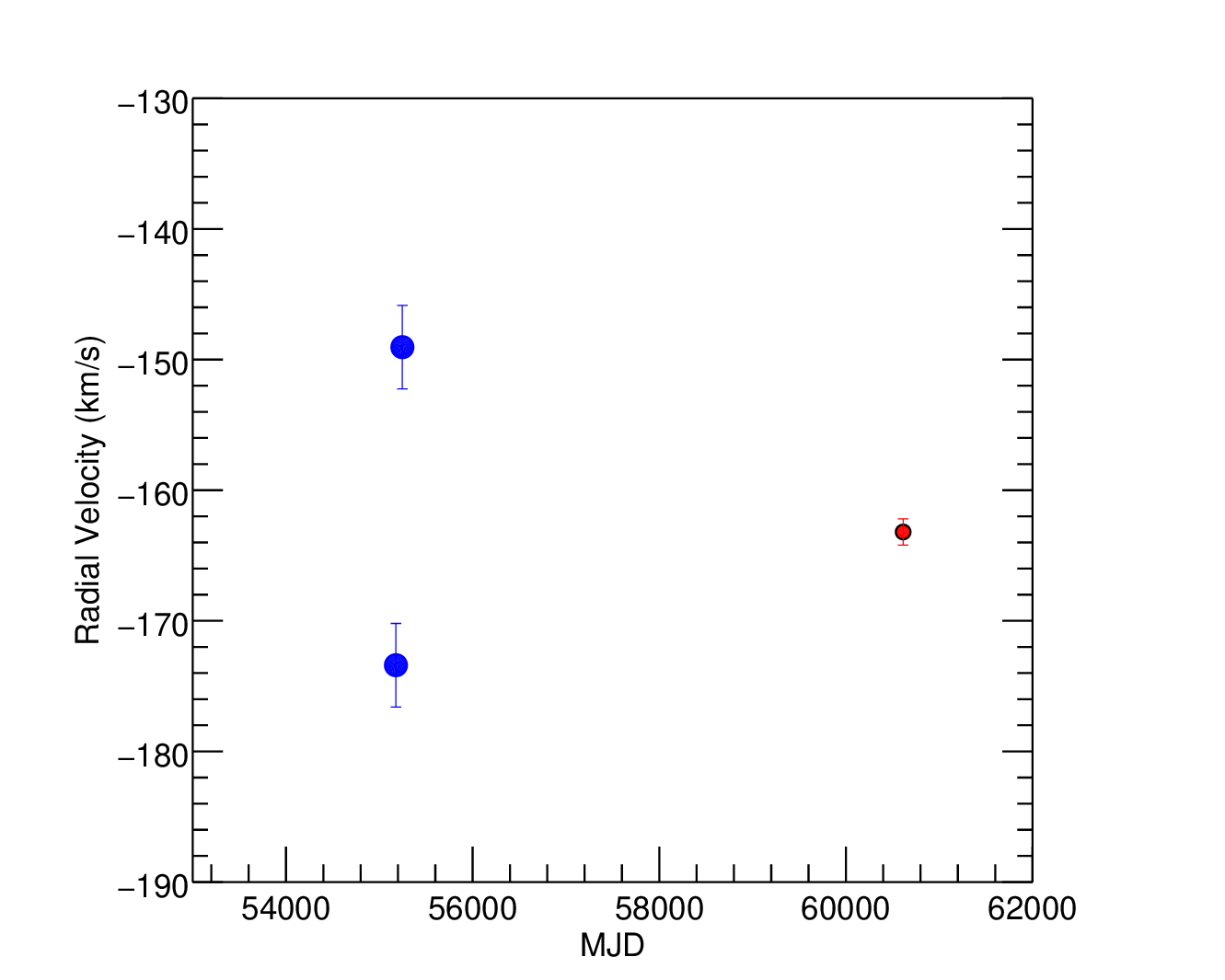} 
 \end{center}
\caption{Radial-velocity variation of the star SDSS~J074238.68+470537.0. Blue symbols represent the results from the SDSS survey. The red symbol represents
the  measure of the radial velocity from our spectrum. }
\label{fig:vr}
\end{figure}

The star SDSS~J074238.68+470537.0 has been identified as a potential EMP stars in the framework of the TOPoS
survey based on the thorough analysis of the SDSS spectra.  Two spectra of this star were taken 
in Dec 2009 and Feb 2010.  Radial velocities computed for this date show a significant difference of the order of 24 km/s, which is significantly larger than the typical reported error of $\simeq 3$ km/s. 
In Fig. \ref{fig:vr} we gather the measured radial velocities for this star. 

\PFR{ The available radial-velocity measurements indicate possible variability, which is suggestive of the presence of a companion. However, the limited number of measurements and their sparse temporal sampling are insufficient to constrain the orbital parameters or to establish the binary nature of the system. Given the important radial-velocity variation over a period of 
a couple of months, further long-term radial-velocity monitoring is therefore indispensable in order to determine the orbital period and eccentricity and to confirm whether SDSS J074238.68+470537.0 is indeed a binary system.}

\begin{figure}[ht!]
 \begin{center}
\includegraphics[width=7.5cm]{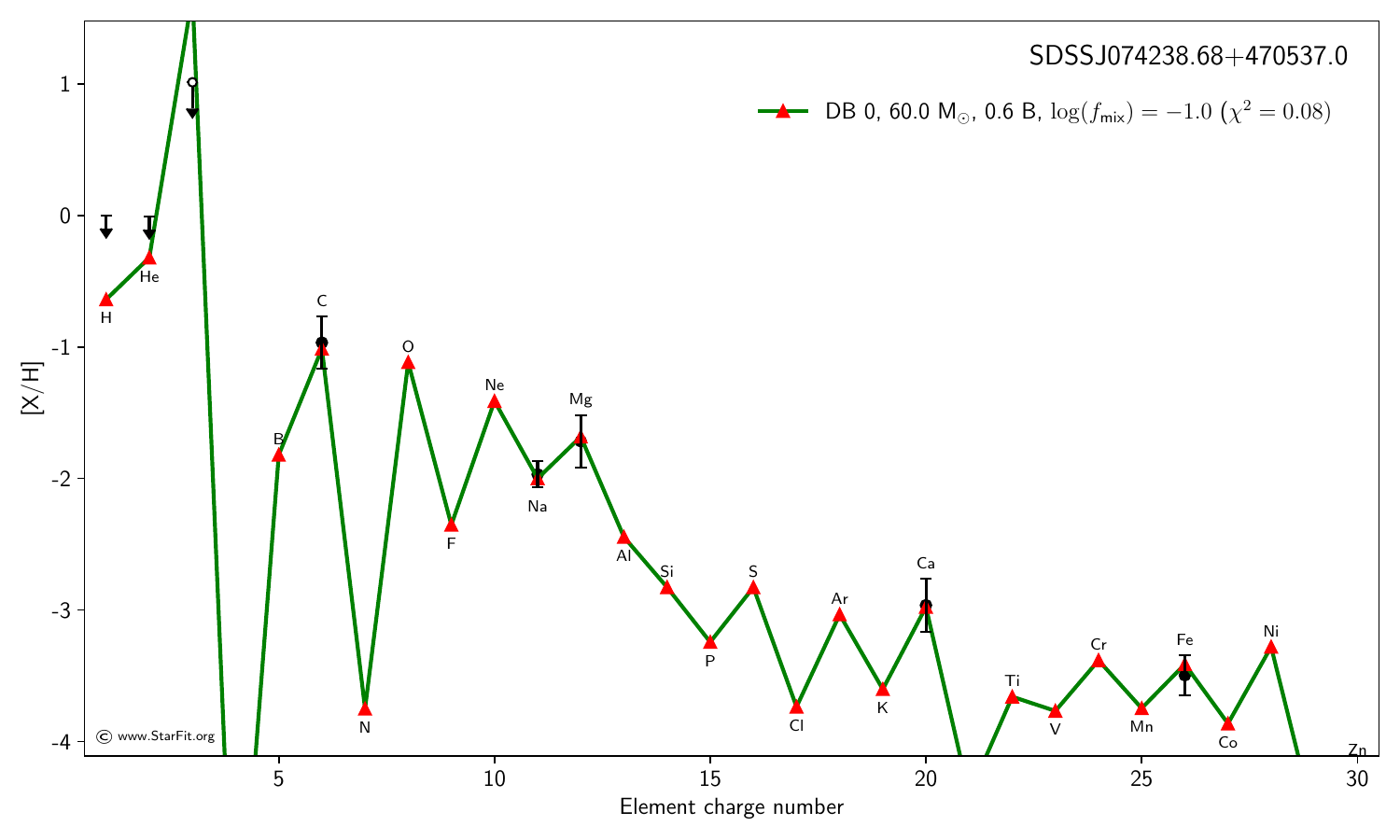} 
 \end{center}
\caption{Fit of nucleosynthesis yields from primordial stars \citep{heger_nucleosynthesis_2010}  to the star SDSS~J074238.68+470537.0 using the STARFIT package (starfit.org). Best fit  is with a 60 $M_{\odot}$ star.}
\label{fig:starfit}
\end{figure}

In Fig. \ref{fig:BaMg} we show the [Mg/Fe] ratio as a function of [Ba/Fe] for the  CEMP stars found in  the SAGA database. 
It is interesting to note the location of the \cemps\ stars, particularly the dwarf stars that show high  [Ba/Fe] ratio  ranging from $\simeq$ 0.8 dex to more than +3 dex. 
Our star stands out with its extremely high [Mg/Fe] ratio, extending the range of the [Mg/Fe] ratio to +1.85 dex.  SDSS~J074238.68+470537.0 shares the same  high [Mg/Fe] ratio as the CEMP dwarf\  HE~1327-2326 \citep{aoki_he_2006}. However, the latter star has a much lower metallicity with [Fe/H]= -5.6 dex. The star HE~0017-4346 \citep{cohen_normal_2013} can also be used as a comparison to SDSS~J074238.68+470537.0. It has a metallicity [Fe/H] = -3.07 dex, slightly higher than our star.
It is a \cemps\ star with [C/Fe]=+2.9 and [Na/Fe]=1.38 and  [Ba/Fe]=1.28 similar to the  [C/Fe]=+2.57, [ Ca/Fe]=1.5 and [Ba/Fe]=1.35 found for our star. However, the [Mg/Fe] found in our star is 1 dex higher than the abundance found for  HE~0017-4346. This large difference could  be due to a different supernova that polluted the gas cloud from which  our star was formed.  
Figure \ref{fig:starfit} shows a comparison of our abundance  results for SDSS~J074238.68+470537.0 with a fit of nucleosynthesis yields from primordial stars \citep{heger_nucleosynthesis_2010}.  As a substantial fraction of Sr and Ba present in the star comes from the  pollution by the ejecta of a past thermally-pulsing  asymptotic giant branch (AGB) companion (now a white dwarf), we did not include these two elements in the fit.
The best fitting supernova out of stars in the 10 $M_{\odot}$ to 100 $M_{\odot}$ range is from a 60 $M_{\odot}$ star. This result must be taken with caution as the fit is based on  only five elements. However, it illustrates the need for high-mass pre-supernovae to explain the high  relative Mg abundance 
found in this star.

The \sit\ in EMP AGB stars is a promising site to be confirmed as the origin of \cemps\ stars because of the neutron sources activated during the thermally pulsing AGB phase.
In particular, \sit\ nucleosynthesis dramatically alters the abundance patterns through Helium-Flash Driven Deep Mixing (He-FDDM; \citealt{Fujimoto2000,Suda2004}) in low-mass, low-metallicity AGB stars.
This phenomenon occurs at the beginning of the thermally pulsing AGB phase through hydrogen ingestion from the envelope into the helium-flash convective zone.
It is also referred to as a dual-shell flash or proton-ingestion episode \citep{Campbell2008,GilPons2018} and can lead to intermediate neutron-capture (\textit{i}-process) nucleosynthesis \citep{Hampel2016,Choplin2021}.
Such events are found in many 1D stellar evolution calculations \citep{Herwig2003,Lau2009,Lugaro2012,Cristallo2016}.

\begin{figure}[ht!]
 \begin{center}
\includegraphics[width=7.5cm]{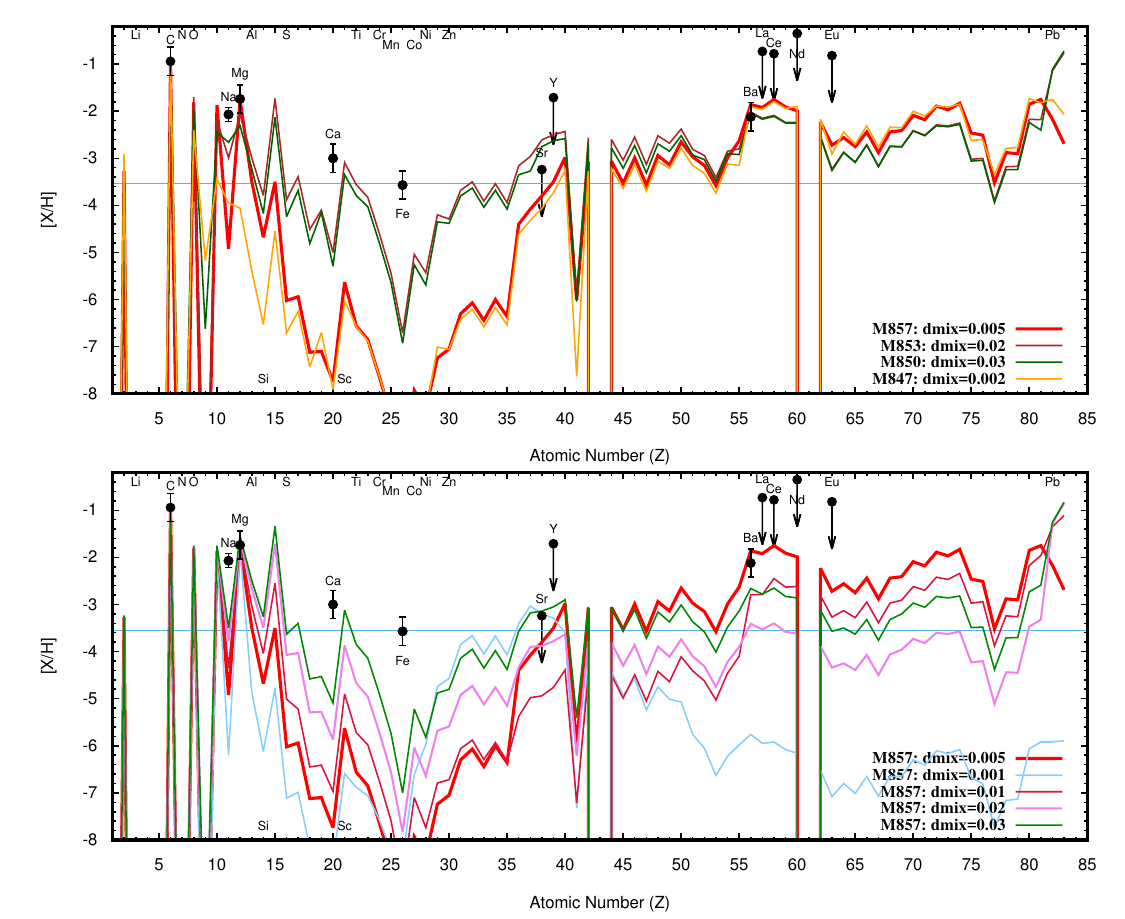} 
 \end{center}
\caption{Fit of \sit\ nucleosynthesis models for EMP AGB stars \citep{Yamada2023} to the observed abundances of \sdsscemp\ using the nuclear-reaction network for the \sit.
Models with very high temperatures were adopted to reproduce the enhanced magnesium abundance.
\suda{The mixing parameter, indicated by \texttt{dmix} in the legend, characterizes the degree of $^{13}{\rm C}$ mixing into the helium-flash convective zone and is defined as the abundance ratio $^{13}{\rm C}/^{12}{\rm C}$.}
The model names given in the legend correspond to the maximum temperature reached during the helium-shell flash. \suda{The top and bottom panels show the dependence of the resulting abundance patterns on temperature and the degree of mixing, respectively.} See the text for details.}
\label{fig:sp}
\end{figure}

Figure~\ref{fig:sp} compares the observed abundances of \sdsscemp\ with nucleosynthesis models including He-FDDM.
We performed \sit\ nucleosynthesis calculations using the code developed by \citet{Yamada2023}, which is based on an extension of the nuclear reaction network for AGB stars by \citet{Nishimura2009}.
To reproduce the high magnesium abundance observed in \sdsscemp, we adopted the M857 models from \citet{Yamada2023}, in which the maximum temperature at the base of the helium-flash convective zone reaches $\log T = 8.57$.
The initial abundances in the model are set such that the carbon mass fraction in the helium convective zone is $X_{\rm C} = 0.1271$.
The abundances of elements heavier than oxygen are assumed to follow scaled-solar abundances with [Fe/H] $= -3.57$, consistent with the metallicity of \sdsscemp.
We explored several sets of parameters describing the initial conditions of the helium-flash convective zone, including the temperature, density, mass, and radius of the core.
The amount of ingested hydrogen is  \PFR{parametrised} by the abundance ratio $^{13}{\rm C}/^{12}{\rm C}$ in the helium convective zone.
The duration of the mixing during the thermal pulse is treated as another free parameter and was fixed at $10^{8}$~s in this study.

\suda{The model dependence on temperature and the amount of mixing is shown in the top and bottom panels of Fig.~\ref{fig:sp}, respectively. In the top panel, the model parameters are chosen to reproduce the barium abundance for models with different maximum temperatures in the helium-flash convective zone. The significant enhancement of magnesium requires a high temperature at the bottom of the helium-flash convective zone, which generally corresponds to a high core mass. However, because He-FDDM occurs during the early phase of the thermally pulsing AGB evolution of low- and intermediate-mass stars, such a high core mass is generally not achieved.
In the bottom panel, the model parameters were chosen to reproduce the magnesium abundance for the M857 models with different values of the mixing parameter. The dependence of the abundance pattern on the amount of mixing is complicated by the production efficiency of \sit\ elements. In particular, the production of heavy elements depends strongly on the neutron exposure as discussed below.
}

We found the best fit with the M857 model and a mixing parameter of $^{13}{\rm C}/^{12}{\rm C} = 0.005$.
\suda{We focused on the magnesium enhancement and the abundance pattern of \sit\ elements within the framework of nucleosynthesis in EMP AGB stars. The abundances of $\alpha$- and iron-group elements are expected to reflect the contribution from supernova nucleosynthesis and were therefore not considered in the fitting, while the calcium abundance is typical of EMP stars.
The sodium abundance of \sdsscemp\ is too high to be reproduced by any of the parameter combinations explored in this study.}
The abundance pattern of the \sit\ elements is reasonably reproduced when the amount of mixing is moderately large.
For smaller amounts of mixing, the \sit\ elements are not sufficiently produced, as illustrated by the M857 model with $^{13}{\rm C}/^{12}{\rm C} = 0.001$.
For larger amounts of mixing, two effects make it difficult to reproduce the observed barium abundance.
One is the higher neutron exposure, which enhances lead production at the expense of the original heavy elements in the helium-flash convective zone.
This effect is seen in the case of $^{13}{\rm C}/^{12}{\rm C} = 0.001$, where barium is only weakly enhanced, while lead is significantly more abundant than in the model with $^{13}{\rm C}/^{12}{\rm C} = 0.0005$.
The other is the efficient production of strontium through neutron-capture nucleosynthesis starting from lighter elements such as oxygen and magnesium \citep{Yamada2023}.
For the M857 models with mixing parameters of $^{13}{\rm C}/^{12}{\rm C} = 0.002$ and $0.003$, the predicted strontium abundances exceed the observed upper limit.

It should be noted that the very high magnesium abundance cannot be reproduced by ordinary AGB models.
The stellar models require temperatures as high as $3.8 \times 10^{8}$~K, which are not achieved in low-mass AGB stars.
This can be seen from the M847 and M850 models in Fig.~\ref{fig:sp}, which reproduce the abundance pattern of the \sit\ elements reasonably well but fail to reproduce the observed magnesium enhancement.
According to the He-FDDM AGB models of \citet{Suda2010}, the maximum temperature at the base of the helium-flash convective shell only reaches $2.97 \times 10^{8}$~K.
This discrepancy may indicate the need for more massive AGB stars capable of triggering He-FDDM, or alternatively a very late He-FDDM event occurring after the carbon-oxygen core has become highly compact.
\suda{For example, \citet{Choplin2022} reported a temperature of $3.5 \times 10^{8}$~K for their $3 M_{\odot}$ model, although they did not find any He-FDDM events in this model.}
This difficulty in simultaneously achieving a sufficiently high temperature and triggering He-FDDM is consistent with the observational fact that \cemps\ stars with enhanced magnesium abundances are extremely rare.

\suda{Our models reach a maximum neutron density of approximately $10^{12}\ {\rm cm}^{-3}$ and do not attain the high neutron densities of $\sim 10^{14}\ {\rm cm}^{-3}$ during the He-FDDM event. This is in contrast to the results of \citet{Choplin2021,Choplin2022}, whose authors found that He-FDDM leads to the so-called {\it i}-process, which is characterised by substantially higher neutron densities. Compared with the \sit, the {\it i}-process generally favours the production of heavier neutron-capture elements and can produce distinctive abundance signatures, such as those involving Xe ($Z=54$) \citep{Choplin2021}. Since the resulting abundance patterns depend on the adopted stellar models and the treatment of mixing, it may be difficult to distinguish between the two processes based solely on the observed abundance pattern.
}

We assume the binary mass-transfer scenario for the origin of \cemps\ stars, in which the observed \cemps\ star is polluted by material transferred from an AGB companion in a binary system.
The abundance pattern of \sdsscemp\ is reasonably reproduced after calibrating the carbon abundance.
This approach is motivated by the idea that most of the carbon observed in \cemps\ stars originates from the binary companion, while the final carbon abundance depends on both the carbon enrichment in the AGB envelope and the efficiency of binary mass transfer.
The final iron abundances shown in Fig.~\ref{fig:sp} correspond to those in the helium-flash convective zone, where most of the iron nuclei are exposed to neutron captures during the \sit.
This does not affect the comparison with the observed abundances because the iron observed in \cemps\ stars mainly originates from the envelope of the observed star itself.

\suda{
We attempt to provide a quantitative estimate for the binary mass-transfer scenario. Our model results can be compared with the observations using the carbon abundance as a reference. The carbon abundance in the helium-flash convective zone is much higher than that in the envelope and than the observed surface carbon abundance. Therefore, the abundances of \sit\ elements relative to carbon are expected to remain approximately unchanged during the mass-transfer and subsequent dilution processes. In addition, the surface carbon abundance of an AGB star does not vary significantly after the He-FDDM event for a given initial mass and metallicity.
}

\suda{
As an example, we adopted the surface carbon abundance of the $1.2,M_{\odot}$ model with [Fe/H] $=-4$ of \citet{Suda2010}, $X_{\rm C}=3.8\times10^{-3}$, as representative of the accreted material. We define the dilution factor $f$ by
\begin{equation}
X_{\rm obs}=(1-f)X_{\rm acc}+fX_{\rm ini},
\end{equation}
where $X_{\rm obs}$ is the observed surface abundance, $X_{\rm acc}$ is the abundance in the accreted material, $X_{\rm ini}$ is the initial abundance of the observed star, and $f$ is the dilution factor. Using the carbon abundance measured for \sdsscemp\ in this study, we obtain $f\sim0.9$. Since the observed \cemps\ star is a dwarf, its convective envelope mass is expected to be of the order of $10^{-3},M_{\odot}$. The required accreted mass is estimated to be $\sim 10^{-4},M_{\odot}$. Thus, only a small amount of transferred material is required to produce the observed surface abundance enhancement.
}

\suda{Our models are subject to theoretical uncertainties associated with the stellar models and the treatment of mixing. The stellar models are based on 1D calculations using standard mixing-length theory \citep{Suda2010} and are therefore subject to uncertainties in the mixing prescription, particularly when the nuclear-burning timescale becomes comparable to the convective turnover timescale. Uncertainties in nuclear reaction rates also affect the resulting abundance patterns. A comprehensive investigation of the sensitivity of our results to all these uncertainties is beyond the scope of this work and will be presented in a forthcoming paper.
}

\section{Discussion and conclusions}
In this article, we present a detailed chemical analysis of three
extremely metal-poor candidates observed with the high-dispersion spectrograph (HDS) at the Subaru telescope.
None of these stars has previously been studied at high-resolution. Their selection was made on the analysis of SDSS spectra 
using the  method described by \citet{ludwig_extremely_2008} to estimate the metallicity of
turn-off stars from low-resolution spectra. 
Among these stars, we discovered a new \cemps\ dwarf star SDSS~J074238.68$+$470537.0, with a metallicity of [Fe/H] = -3.57 dex.
This star exhibits not only a high carbon abundance (A(C)=7.5) and  a high [Ba/Fe] with [Ba/Fe]= 1.35 dex, it also has a very high 
magnesium enrichment, with a ratio o [Mg/Fe] = 1.85.
We were able to determine the abundances of some elements (C, Mg, Ca, Sr, and Ba) in the majority of these stars.
We measured a lithium abundance of A(Li) = 2.05 dex in the star SDSS~J012630.30$+$073029.3 (metallicity [Fe/H] = -3.02, where the lithium plateau meltdown
begins  to appear). Two stars of the sample show abundance ratios that are typical of metal-poor stars 
in the metallicity range $-3.57$ dex $\le$ [Fe/H] $\le -3.05$ dex, whereas the last one is a \cemps\ star with a high [Mg/Fe].

The fact that SDSS~J012630.30$+$073029.3  and SDSS~J074238.68$+$470537.0 are candidate members of GSE
is interesting because it suggests that the progenitor of GSE started forming stars
from primordial gas and not pre-enriched gas. A purely kinematical selection of GSE, as used 
here, is prone to including a contamination of the order of 20\% \citep{bonifacio_topos_2021}; 
however, little is known about the abundance patterns of GSE in the EMP regime, so we cannot
rely on chemistry to confirm or refute the association of these stars with GSE. If they are indeed 
members of GSE, the fact that their abundance pattern is similar to that of other halo stars
in the same metallicity range would suggest that the GSE progenitor and the Milky Way
had a similar chemical evolution in their early stages. 

The star SDSS~J214340.08$-$002835.5 increases the number of EMP stars found in disc orbits \citep{sestito_pristine_2020,di_matteo_reviving_2020, dovgal_probing_2024}.
\citet{fernandez-alvar_pristine_2021} found evidence of an EMP disc; however, they considered
their sample of stars in this component too small to provide compelling evidence of its existence. Such an existence is, however, supported by the observations at high spectral resolution.

\begin{acknowledgements}
We are grateful to L. Monaco for many interesting discussions on Galactic dynamics and for his guidance in the use of {\tt galpy}. TS acknowledges the support of Grants-in-Aid for Scientific Research ( KAKENHI) (JP22K03688, JP25HP8011, JP26K07156, JP26HP8010) from the Japan Society for the Promotion of Science (JSPS). TS and WA also acknowledge support from JSPS KAKENHI (JP25K01046). PF acknowledges support from the AIPS and the  AI-GGS of the Paris-Meudon Observatory.
\end{acknowledgements}





%


   \bibliographystyle{aa} 

\bibliography{CEMP_Subaru} 
%




\end{document}